\pdfoutput=1

\documentclass[11pt]{article}
\usepackage{acl}

\usepackage{times}
\usepackage{latexsym}
\usepackage{comment}
\usepackage[T1]{fontenc}
\usepackage[utf8]{inputenc}
\usepackage{microtype}
\usepackage{inconsolata}
\usepackage{graphicx}

\title{Deep Learning-based Bug Triage System}
\author{Sourabh Pal, OLAS Team, INRIA/University of Bologna, sourabh.pal@inria.fr}

\begin{document}
\maketitle

\begin{abstract}
Effective bug triage is crucial for streamlining the software development lifecycle by accurately categorizing and assigning reported software defects. In this paper, we propose an automated bug triage system built upon the pre-trained RoBERTa-base transformer architecture. By leveraging deep contextual representations, our approach efficiently classifies incoming bug reports to optimize assignment. Experimental evaluation demonstrates that the proposed system achieves a strong bug identification accuracy of 0.90 within just five training epochs. These findings highlight the efficiency and high performance of fine-tuned transformer models for practical software engineering automation.
\end{abstract}

\section{Introduction}
Software development is a very complex and challenging process. It includes the five steps namely, analysis, design, implementation, testing, and maintenance \cite{SoftwareDevelopmentLifeCycle}. Software Maintenance requires nearly 40-80\% of the overall software production costs and there are plenty of tools designed to help developers \cite{SoftwareMaintenance01, SoftwareMaintenance02}. The developers mostly analyze the bug reports and spend significant time understanding the specific parts of the code responsible for producing the bugs. It consumes over 30\% of the time of software maintenance \cite{SoftwareMaintenance03}. 

The user of the software product mainly submits the issue in case of failure of the software system. While the user reports the bug, the bug information is stored in the issue tracking system. Large software systems consist of different products and components. It could be convenient for the developers to identify the bug-producing code if the developer identifies the component responsible for producing that specific bug. Conventionally, the user provides the information of the product and component while submitting the bug reports to the issue tracking system. In this research, we conduct the following two experiments:

\begin{itemize}
    \item \textbf{Identify Component:} The easy and fast identification of the buggy component for the new coming bugs can reduce the bug triage resources. Conventionally, the developers or experts manually read the bug reports and assign the bugs to specific components for further assignment to the developer responsible for resolving the bug. The manual process of component identification is time-consuming. In this research, we automatically identify the component for the new upcoming bugs.  
    \item \textbf{Buggy:} It is not always possible that the provided bug reports from the user are always buggy. It could be duplicate bugs, invalid bugs, or already resolved. Therefore, the automatic identification of the actual bug can reduce the utilization of the bug triage resources.  
\end{itemize}

\section{Dataset Description}
Bugzilla \cite{buzilla} is a web-based issue-tracking system that allows developers to track and resolve the bugs of software products. In an issue-tracking system, the users usually raise bugs, and based on the bugs report the the software practitioners triage the bugs. We have collected only the \textbf{Client Software} type of software product from the web repository. This type contains sixteen different products namely, Calendar, Cloud Services, Data Platform and Tools, Fenix, Firefox, Firefox for iOS, Firefox Private Network, Focus, Focus-iOS, Mozilla Localizations, Mozilla VPN, Other Applications, Pocket, SeaMonkey, Thunderbird, and Web Compatibility.

We extracted the following information from the issue tracking system:
\begin{itemize}
    \item \textbf{BugSummary:} It describes the issue that arises in the software system.
    \item \textbf{Product:} It defines the which software software product affect due to bug.
    \item \textbf{Component:} It defines the which component of the software software product affect due to bug.
    \item \textbf{Resolution:} It defines the current status of the bug. It consists of ten different status as follows:
        \begin{itemize}
            \item \textbf{FIXED: } It indicates that a particular commit has fixed the bug. 
            \item \textbf{INVALID: } It indicates that something about this bug report is invalid or does not compute.
            \item \textbf{WONTFIX: } It indicates that this may be a bug, but we aren't planning to fix it for some reason because it would not affect users.
            \item \textbf{INACTIVE: } It indicates that a bug report is no longer being worked on or monitored for various reasons such as being closed, and so on.     
            \item \textbf{DUPLICATE: } It indicates that the bug is a duplicate of another bug already in the system.
            \item \textbf{WORKSFORME: } It defines that the tester has tested the bug which could reproduce it in an older version, but can't anymore in a current version. 
            \item \textbf{INCOMPLETE: } It defines that the bug reports do not provide enough information to be tackled.
            \item \textbf{EXPIRED: } The issue bugs are expired because the issues are probably already solved.
            \item \textbf{MOVED: } The Bugzilla is not the proper place to report this bug. So, it should move to the other issue tracking system.
        \end{itemize}
\end{itemize}

We extracted 69431 bug reports \cite{bugDataset} from the Bugzilla issue tracking system.

\section{Architecture Overview}
We represent an automated approach to identify the affected component and classify the software bug for the upcoming bugs. Our approach consists of three steps, namely \textbf{Data Collection}, \textbf{Data Cleaning}, and \textbf{Classification}. We discuss each steps of our approach as follows.

\subsection{Data Collection}
Software bug reports stored in the issue tracking system. We need to extract the relevant information from this issue tracking system for further automation of the bug triage activities. In our case study, we extracted the Bug Summary, Component and Resolution from the Bugzilla issue tracking system to store it the the database for further analysis of our research.  

\begin{figure}[t]
  \includegraphics[height=15.0cm, width=\columnwidth]{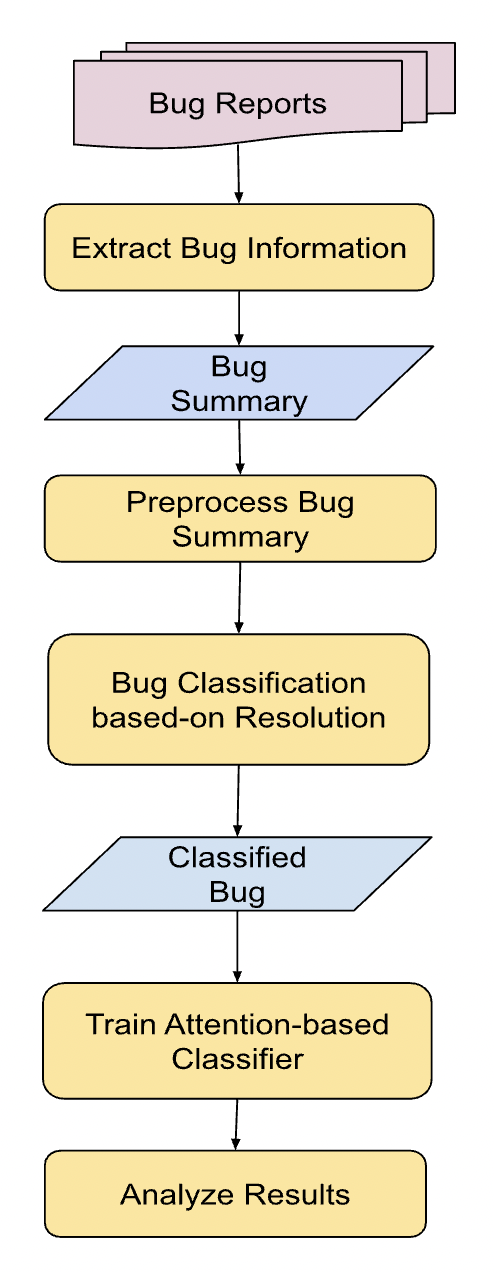}
  \caption{A Schematic representation of our proposed approach.}
  \label{fig:model}
\end{figure}

\subsection{Data Cleaning}
The bug description or summary usually available in the form of natural language. Therefore, we apply the standard natural language processing techniques (NLP) to clean or preprocess our data for further analysis. We applied the three approaches to preprocess our NLP based bug summary, namely Special Character Removal, Stop Word Removal, and Lemmatization. We discuss this approaches as follows.
\begin{itemize}
    \item \textbf{Special Character Removal: } In this step, we remove all punctuation symbols (., !, etc.) are removed, and the alphanumeric symbols (‘*’, ‘@’, ‘\#’, etc.) appear in the in the sentence are removed and replace with empty spaces.
    
    \item \textbf{Stop Word Elimination: } The stop words includes the article, helping verb, modal verb, pronoun and so on. This are the most frequently used words in the natural language. Therefore, they does not contribute to the modeling of the natural language based tasks \cite{plisson2004rule}. So, the Stop Word Elimination may improve the performance of the model.    
    \item \textbf{Lemmatization: } We used the lemmatization techniques to convert the words of the bug summary into its canonical form. For example, the word communicating can be lemmatized into its canonical form communicate. The main intuition behind this step is that it groups all the related words with the same root word. This step in NLP makes it easier for prediction and generation tasks.
\end{itemize}

\subsection{Classification}
Our approach uses deep learning techniques. We hypothesize that our technique should exhibit improved learning from the contextual features of the bug resolution summary report due to the use of the attention mechanism. This should lead to boosting the performance of the classifier. We schematically represent our proposed approach in \ref{fig:model}.

\section{Experimental Setup}
We conducted our experiment on a single GPU with a Tesla T4 environment. In this paper, we analyze two research tasks namely, the identification of the corresponding buggy component for the new upcoming bug and classify whether the newly assigned bug is buggy or non-buggy. We used the Bug Summary form the the dataset for the modeling of our model. Additionally, we use the component metrics for the labeling of the component to identify the suitable component of the new bugs, whereas we use the Resolution metrics for labeling for the identification of the bugginess of the new bugs. We labelled the bug as buggy based on the resolution, if the resolution is FIXED, we labelled it as 1 which is buggy. Otherwise, we labelled it as 0 which is non-buggy.

We use the Transformer-based language model RoBERTa \cite{liu2019roberta} for our experiment. In the case of the training and test datasets, we have chosen randomly 90 percent of the data for the training and 10 percent of data for the test of our model.

\section{Results and Analysis}
We use the RoBERTa base transformer with only five epochs due to the lack of available infrastructure. In the experiment, we will analyze the training loss and the validation loss of our model to verify the performance of our model. The training loss (red line) and validation loss (blue line) of component classification and buggy classification are provided in Figure \ref{fig:exp01} and Figure \ref{fig:exp03} respectively. Furthermore, we will measure the accuracy of our model for both research questions. The results analysis of our research questions are following:

\begin{itemize}
    \item \textbf{Component Identification: } This is the multiclass classification because software products generally consist of multiple components. The identification of the right component for the newly assigned software bugs could reduce the effort and resources of the bug triage approaches. In our collected datasets, we have a total of 325 unique components. The training loss of our model is decreasing over the Epochs in Figure \ref{fig:exp01} which indicates that our model is learning to fit the training data better with each epoch. Additionally, the validation loss of our model is also decreasing with increasing the number of epochs. It indicates that our model is training well for the prediction of unseen data. The accuracy of the proposed model of the software component identification is increasing over time with increasing epochs. We have active about 0.7 accuracy with only five epochs in our experimentation in Figure \ref{fig:exp02}. Therefore, our proposed model of component identification provides an opportunity to automatize the process of component identification in the software bug triage system.
    
    \begin{figure}[t]
    \includegraphics[height=5.0cm, width=\columnwidth]{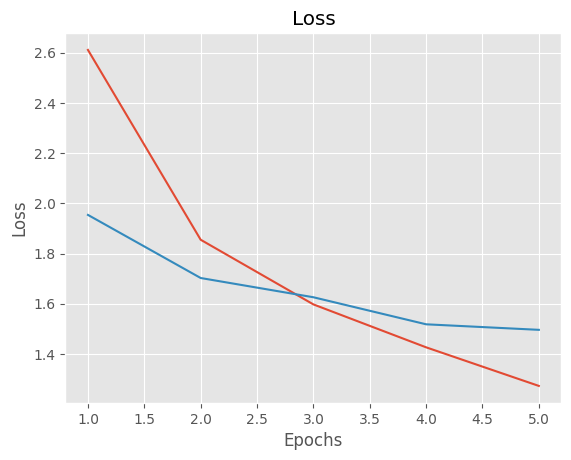}
    \caption{Training Loss and Validation of Component Classification.}
    \label{fig:exp01}
    \end{figure}
    
    \begin{figure}[t]
    \includegraphics[height=5.0cm, width=\columnwidth]{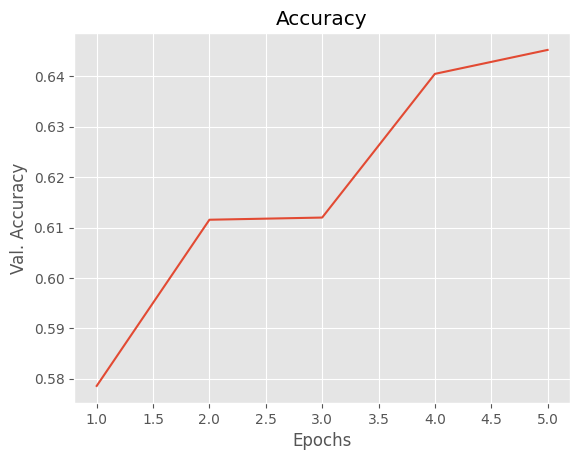}
    \caption{Accuracy of Component Classification.}
    \label{fig:exp02}
    \end{figure}
    
    \item \textbf{Bug Identification: } This is the binary classification where our model will identify whether the newly assigned bug is buggy. The early identification of the bugginess of the newly assigned bug could reduce the effort and resources of the bug triage approaches. The training loss of our model is decreasing over the Epochs in Figure \ref{fig:exp03} which indicates that our model is learning to fit the training data better with each epoch. Additionally, the validation loss of our model is also increasing with increasing the number of epochs. It indicates that our model is suffering from the overfitting problem due to the noisy data. Moreover, we have active accuracy over 0.9 with only five epochs in our experimentation in Figure \ref{fig:exp04}. Therefore, our proposed model is performing well with the given datasets with slightly overfitting issues.
    
    \begin{figure}[t]
    \includegraphics[height=5.0cm, width=\columnwidth]{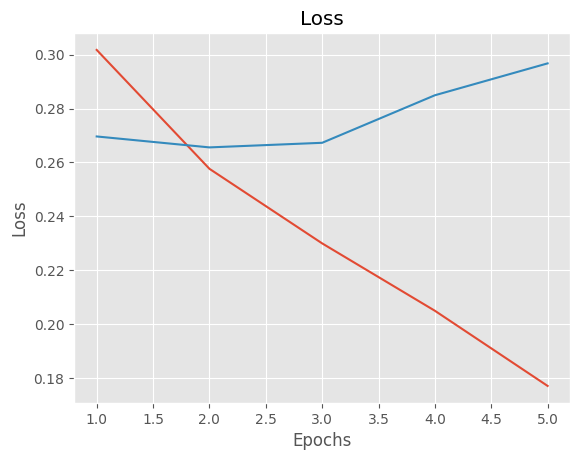}
    \caption{Training Loss and Validation of Bug Classification.}
    \label{fig:exp03}
    \end{figure}

    \begin{figure}[t]
    \includegraphics[height=5.0cm, width=\columnwidth]{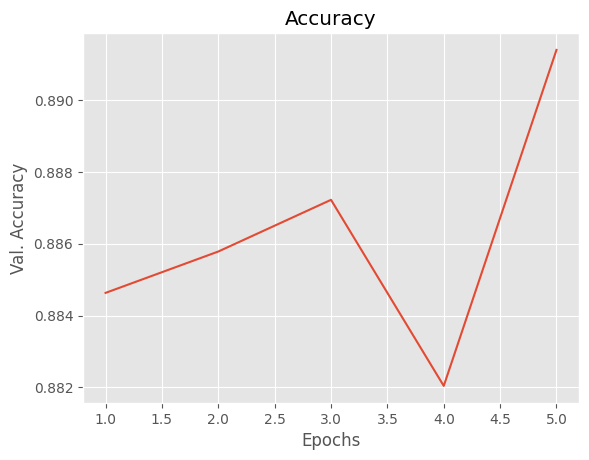}
    \caption{Accuracy of Bug Classification.}
    \label{fig:exp04}
    \end{figure}
    
\end{itemize}

\section{Conclusion}

In this paper, we proposed an automated bug triage system utilizing the fine-tuned RoBERTa-base transformer model to enhance the efficiency of bug categorization and assignment in software development. By capturing deep contextual semantics from bug reports, our system demonstrates remarkable effectiveness, achieving an identification accuracy of 0.90 within only five training epochs. These findings confirm that transformer-based architectures can significantly reduce the manual overhead associated with defect classification while maintaining high precision. 

In future work, we plan to extend this framework by evaluating it on larger and more heterogeneous multi-project datasets, incorporating severity prediction, and exploring lightweight or distilled transformer models to improve runtime efficiency and lower deployment costs.

\bibliography{custom}

\end{document}